\documentclass{appolbZ60}
\usepackage{amsmath,amssymb,float,graphicx,doi}
\begin{document}
\title{The muon abundance in the primordial Universe%
\thanks{Prepared for the 60th anniversary of the Cracow School of Theoretical Physics}%
}
\author{Jan Rafelski and Cheng Tao Yang
\address{Department of Physics, The University of Arizona, Tucson, AZ 85721}
}
\maketitle\vskip -0.6cm
\centerline{March 2021}
\begin{abstract}
Muon abundance is required for the understanding of several fundamental questions regarding properties of the primordial Universe. In this paper we evaluate the production and decay rates of muons in the cosmic plasma as a function of temperature. This allows us to determine when exactly the muon abundance disappears. When the Universe cools below the temperature $kT_\mathrm{disappear}\approx 4.135$ MeV the muon decay rate overwhelms production rates and muons vanish quasi-instantaneously from the Universe. Interestingly, we show that at $T_\mathrm{disappear}$ the muon number is nearly equal to baryon abundance. 
\end{abstract}

\section{Introduction}
Our interest in strangeness flavor freeze-out in the early Universe requires the understanding of the abundance of muons in the early Universe: The specific question needing an answer is the temperature at which muons remain in abundance (chemical) equilibrium established predominantly by electromagnetic and weak interaction processes, allowing detailed-balance back-reactions to influence strangeness abundance.

We recall that the strangeness decay often proceeds into muons, energy thresholds permitting, as the charged kaons K$^\pm$ demonstrate with a 63\% branching into $\mu+\bar \nu_\mu$. Should muons fall out of thermal abundance equilibrium this would directly impact the detailed balance back-reaction processes. Another, indirect influence on strangeness in the early Universe arises through the nearly exclusive decay of charged pions into $\mu+\bar \nu_\mu$. Without muon chemical abundance equilibrium this back reaction stops too, impacting pions and thus all other hadronic particles in the Universe. 

Here it is important to remember that another relatively strong channel for producing pions thermally in the Universe is the fusion process $\gamma+\gamma\to \pi^0$ followed by a strong interaction charge exchange~\cite{Kuznetsova:2010pi}. While pion production via this chain of reactions assures thermal pion abundance down to temperatures that could be numerically explored, the muon chemical abundance equilibrium becomes \lq leaky\rq\ when in some reactions the detailed balance is broken. Detailed balance reactions for pions and thus exact pion chemical equilibrium require muons to remain in thermal abundance equilibrium.

Another reason to be interested in muon abundance is that these particles could help in the model building of a Universe symmetric in baryon number, where baryons and antibaryons are separated into space-time domains. Other exotic Universe models can be imagined in such situations including strangelet formation. We do not discuss these possibilities here, focusing instead exclusively on learning about muon abundance disappearance in the early Universe. This is done here in preparation for the comprehensive report on strangeness abundance~\cite{Rafelski:2020pdi}.

To establish the relevance of the questions considered we first explore the different and relevant muon $\mu^\pm$ production and disappearance rates in the cosmic plasma, section~\ref{sec:prodmudec}. In section~\ref{sec:abund} we characterize the chemical equilibrium abundance of muons in the early Universe, normalized by the baryon abundance, considering the here relevant temperature domain $6\ge T\ge 3$\,MeV. We employ in this consideration the conservation of baryon number and entropy during the Universe evolution at temperatures below $kT<10$\,MeV. After a summary of our results, we add in the Appendix some personal reminiscences about our long-standing connection to the Krak\.ow school of theoretical physics.
 
All natural constants and information about elementary particles and the Universe used in this report were obtained from the {\it Particle Physics Booklet}~\cite{Tanabashi:2018oca}. 


\section{Production and decay of $\mu^\pm$ in the early Universe}\label{sec:prodmudec}
\subsection{Elementary processes} 
In the early Universe in the cosmic plasma muons of mass $E=m_\mu c^2=105.66$\,MeV can be produced by the interaction processes
\begin{align} 
&\gamma+\gamma\longrightarrow\mu^++\mu^-,\qquad & e^++e^-\longrightarrow \mu^++\mu^-\;,\\
&\pi^-\longrightarrow\mu^-+\bar{\nu}_\mu,\qquad & \pi^+\longrightarrow\mu^++\nu_\mu\;.
\end{align}
The back reaction for all above processes is in detailed balance, provided all particles shown on the right-hand side (RHS) exist in chemical abundance equilibrium in the Universe. 

However, all produced muons can decay 
\begin{equation}
\mu^-\rightarrow\nu_\mu+e^-+\bar{\nu}_e,\qquad \mu^+\rightarrow\bar{\nu}_\mu+e^++\nu_e\;.
\end{equation} 
We thus must establish the range of temperature in which production processes exceed in speed the decay process: We recall the empty space (no plasma) at rest life-time of pions $\tau_\pi=2.6033\times10^{-8}$\;s, and that of muons $\tau_{\mu}=2.197 \times 10^{-6}$\;s. 
 
\subsection{Reaction rates} 
\subsubsection{In plasma decay rate}
The temperature range of interest to this investigation is in the time era of the Universe when $m_\mu c^2\gg T$. Thus it is appropriate to study massive particles, here muons (and pions), considering the Boltzmann limit of their Fermi (and Bose) quantum distributions. The thermal decay rate per volume and time in the Boltzmann limit is~\cite{Kuznetsova:2008jt}
\begin{align}
&R_\mathrm{decay} =\frac{g}{2\pi^2}\left(\frac{T^3}{\tau }\right)\left(\frac{m}{T}\right)^2K_1(m/T)\;, 
\end{align}
where the particle degeneracy $g$, mass $m$, and lifespan $\tau$ are given.

This thermal decay rate accounts for the thermal Boltzmann limit density of particles in chemical abundance equilibrium and the effect of time dilation present when particles are in thermal motion with respect to observer at rest in the local reference frame. The effects of Fermi-blocking or Boson-stimulated emission have been neglected when taking the Boltzmann limit.

\subsubsection{In plasma production rate}
The scattering angle averaged thermal reaction rate per volume for the reaction $a\overline{a}\rightarrow b\overline{b}$ in the Boltzmann approximation is given by \cite{Letessier:2002gp}
\begin{align}\label{pairR}
R_{a\overline{a}\rightarrow b\overline{b}}=\frac{g_ag_{\overline{a}}}{1+I}\frac{T}{32\pi^4}\int_{s_{th}}^\infty ds\frac{s(s-4m^2_a)}{\sqrt{s}}\sigma_{a\overline{a}\rightarrow b\overline{b}} K_1(\sqrt{s}/T),
\end{align}
where $s_{th}$ is the threshold energy for the reaction, $\sigma_{a\overline{a}\rightarrow b\overline{b}}$ is the cross section for the given reaction, and we introduce the factor $1/(1+I)$ to avoid the double counting of indistinguishable pairs of particles. We have $I=1$ for an identical pair and $I=0$ for a distinguishable pair.

The invariant matrix element (squared) for the reactions $e^++e^-\to\mu^++\mu^-$ and $\gamma+\gamma\to\mu^++\mu^-$, are, respectively~\cite{Kuznetsova:2008jt}
\begin{align}\label{Mee}
|M_{e\bar e\to\mu\bar\mu}|^2=&32\pi^2\alpha^2\frac{(m_\mu^2-t)^2+(m_\mu^2-u)^2+2m_\mu^2s}{s^2},\quad m_\mu\gg m_e\;,\\[0.2cm]
|M_{\gamma\gamma\to\mu\bar\mu}|^2=&32\pi^2\alpha^2\bigg[\left(\frac{m_\mu^2-u}{m_\mu^2-t}+\frac{m_\mu^2-t}{m_\mu^2-u}\right)+4\left(\frac{m_\mu^2}{m_\mu^2-t}+\frac{m_\mu^2}{m^2_\mu-u}\right)\\[0.1cm] \nonumber
&\hspace{1cm}-4\left(\frac{m_\mu^2}{m^2_\mu-t}+\frac{m^2_\mu}{m^2_\mu-u}\right)^2\bigg]\;,\label{Mgg}
\end{align}
 where $s, t, u$ are the Mandelstam variables
\begin{align}
&s=(p_1+p_2)^2=(p_3+p_4)^2,\quad
&t=(p_3-p_1)^2=(p_2-p_4)^2,\\\nonumber
&u=(p_3-p_2)^2=(p_1-p_4)^2,\quad
&s+t+u=2m^2.
\end{align}
 
The cross section required in Eq.\,(\ref{pairR}) can be obtained by integrating the matrix elements Eq.\,(\ref{Mee}) and Eq.\,(\ref{Mgg}) over the Mandelstam variable $t$ 
\begin{align}
\sigma_{e\bar e\to\mu\bar\mu}&=\frac{1}{16\pi s(s-4m^2_e))}
\int^{m^2_\mu-s/2\left(1-\sqrt{1-4m^2_\mu/s}\right)}_{m^2_\mu-s/2\left(1+\sqrt{1-4m^2_\mu/s}\right)} |M_{e\bar e\to\mu\bar\mu}|^2\,dt \notag\\[0.5cm]
&=\frac{64\pi\alpha^2}{48}\left(\frac{1+2m^2_\mu/s}{s-4m_e^2}\right)\sqrt{1-\frac{4m^2_\mu}{s}},\\[0,3cm]
\sigma_{\gamma\gamma\to\mu\bar\mu}&=\frac{\pi}{2}\left(\frac{\alpha}{m_\mu}\right)^2(1-\beta^2)\left[(3-\beta^4)\ln\frac{1+\beta}{1-\beta}-2\beta(2-\beta^2)\right]\,,\\[0.4cm] \nonumber
\beta&=\sqrt{1-4m^2_\mu/s}
\end{align}
Inserting these previously known cross sections (for $\gamma\gamma$ see Ref.\,\cite{Gould:1967zzb}, $e\bar e$ is in textbooks) into Eq.\,(\ref{pairR}) we obtain the two thermal production rates.
 
\subsection{Muon disappearance temperature}
\begin{figure}
\begin{center}
\includegraphics[width=10.5cm]{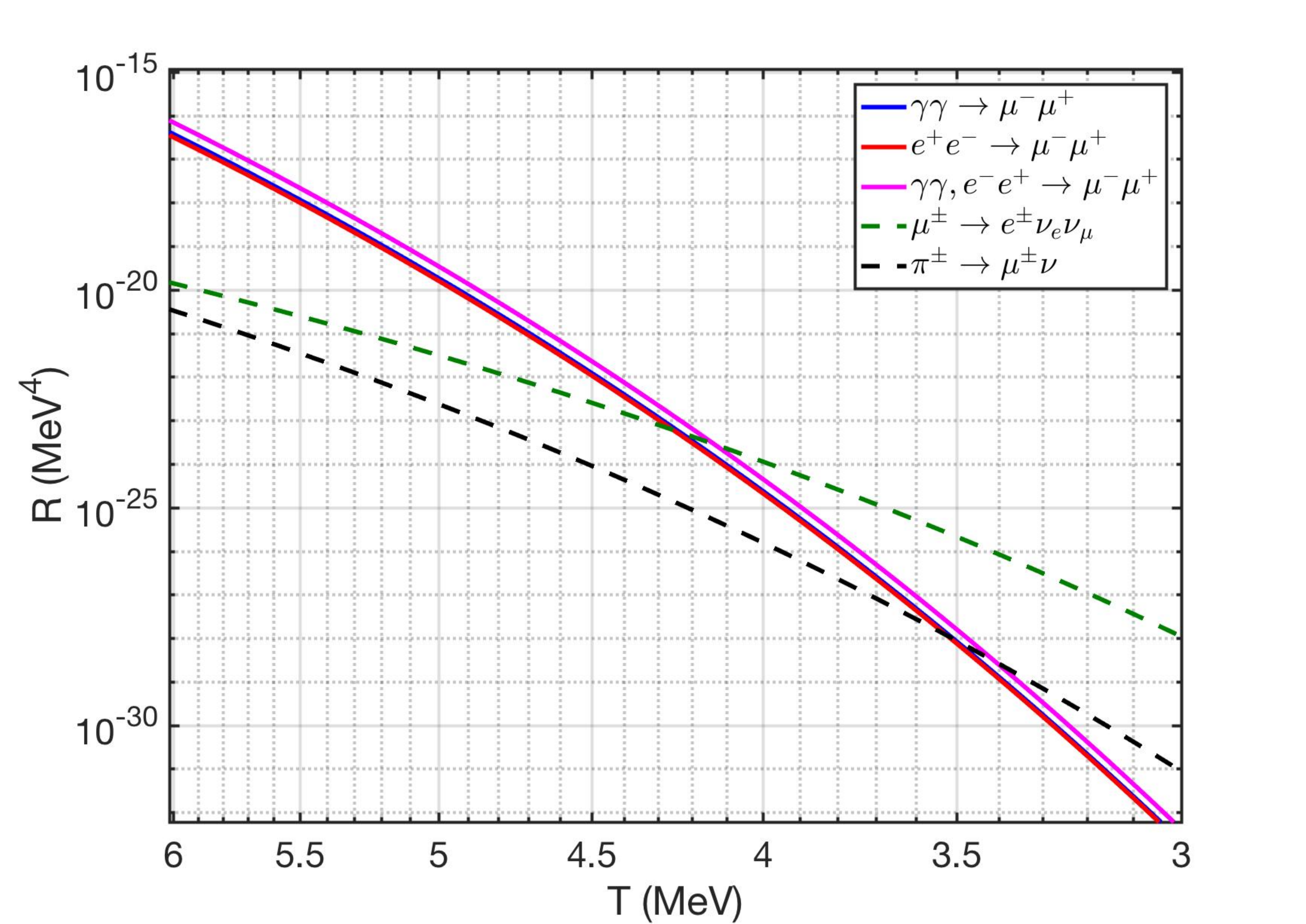}
\caption{\label{MuonRatenew_fig}Thermal invariant reaction rates R per volume and time for different reactions (see box) as a function of temperature $T\in \{6,3\}$\,MeV.}
\end{center}
\end{figure}

In Fig.~\ref{MuonRatenew_fig} we show relevant invariant thermal reaction rates per volume and time as a function of temperature $T$. The red and blue solid lines represent the muon production $e\bar e \rightarrow \mu\bar\mu $ and $\gamma \gamma\rightarrow\mu\bar\mu $ respectively. The purple solid line represent the summed total production rate. The green and black dashed lines are for the decay of $\mu^\pm$ and $\pi^\pm$, respectively.

As the temperature decreases in the expanding Universe, the initially dominant production rates become smaller faster and cross the decay rates. Muon abundance disappears as soon as the muon decay rate is faster than the total production rate -- we include the decay of the pion which produces a muon to show that this channel of muon production does not contribute significantly ever as it is always overwhelmed by the muon decay rate. The difference in mass between these two particles in this temperature range overwhelms the greater speed of pion decay. 

Considering the slow speed of the Universe expansion the muon disappearance is sudden; the muon abundance thus disappears as soon as a decay rate crosses the dominant production rate. Specifically, the muon decay rate (dashed green line) dominates the total muon production rate (purple solid line) below $T_\mathrm{disappear}\approx 4.135$ MeV. 

\section{The number density of $\mu^\pm$}\label{sec:abund}
We now aim to relate the muon abundance to baryon abundance. We know in the present epoch (subscript $t_0$) the baryon $B$ to photon $N_\gamma$ ratio
$$
\frac{B}{N_\gamma} =\left.\frac{n_\mathrm{B}}{n_\gamma}\right|_{t_0}\simeq 0.61\times10^{-9}\;,
$$
where baryon density $n_\mathrm{B}$ and photon density $n_\gamma$ is introduced -- we use the central value from table 2 in Ref.\,\cite{Tanabashi:2018oca} noting $\pm 5\%$ error. We convert this measured value to a ratio of baryon number per entropy of photons in the present epoch by using the universal entropy per particle for a massless boson relation $(s/n)_{\mathrm{boson}}\approx 3.6$~\cite{Letessier:2002gp} to obtain
\begin{align}\label{entropy_per_baryon}
\left.\frac{n_\mathrm{B}}{s_\gamma}\right|_{t_0}=
\left.\frac{n_\mathrm{B}}{n_\gamma}\right|_{t_0}\frac{1}{s_\gamma/n_\gamma}\simeq 1.7\times10^{-10}\;.
\end{align}

The entropy density $s$ can be characterized using the analytic form for massless bosons introducing $g^s_\ast$, the number of \lq entropic\rq\ degrees of freedom
\begin{align}\label{entrop}
s=\frac{2\pi^2}{45}g^s_\ast T^3\;.
\end{align}
For temperature $10 >T>3 $\,MeV, the massless photons, nearly relativistic electron/positrons, and practically massless neutrinos contribute to the degrees of freedom $g^s_\ast$. Therefore, allowing for factor $7/8$ for Fermions as compared to Bosons, we have $g^s_\ast=2+7/8(4+3\times 2)=10.75$, showing each contribution mentioned in turn.

Below $T=3$\,MeV neutrinos decouple, and free stream while electrons and positrons annihilate into photons. This process reheats the photon gas, transferring practically all $e\bar e$-entropy there. This means that the fraction of entropy neutrinos carry $s_\nu/s=21/43=0.49$ stands forever apart from the entropy fraction photons and electrons carry: $s_{\gamma e}/s=22/43=0.51$. This also implies $s_\nu/s_{\gamma e}=21/22=0.95$, a ratio which will be very handy soon. 

Due to very small reheating to neutrinos by annihilating electrons this ratio is about 0.97 according to current theoretical models~\cite{Birrell:2014uka}. We note that the photon entropy $s_\gamma$ in the current epoch we used in Eq.\,(\ref{entropy_per_baryon}) is actually identical to the entropy $s_{\gamma e}$ of the combined photons and annihilated $e\bar e$ pairs applicable to the entire Universe evolution history of relevance here. 

We now exploit that the ratio of baryon number to entropy remains unchanged in the Universe evolution, at least since the muon freeze-out epoch we are considering here. This is so since the baryon number is conserved, and the Universe is expanding adiabatically. We will now take advantage of this feature, apportioning the entropy among $\gamma e$ and neutrino $\nu$ gases to evaluate the muon $\mu^\pm$ to baryon density $n_\mathrm{B}$ ratio. To achieve this we consider 
\begin{align}\label{nmuperb}
\frac{n_{\mu^\pm}}{n_\mathrm{B}}=\frac{n_{\mu^\pm}}{s}\frac{s}{n_\mathrm{B}}=
\frac{n_{\mu^\pm}}{s}\left(\frac{s_{\gamma e}(1+s_\nu/s_{\gamma e})}{n_\mathrm{B}}\right)_{\!t_0}\simeq
\frac{n_{\mu^\pm}}{s}\;\frac{1.97}{1.7\times10^{-10}}\;,
\end{align}
where we used the standard cosmology assumption that $s/n_\mathrm{B}$ remains constant, and that the entropy in the Universe is (vastly) dominated by the two fractions, $\gamma e$ and neutrino $\nu$. 
 
We have for the entropy density $s$ our expression Eq.\,(\ref{entrop}). The number density for nonrelativistic $\mu^\pm$ is
\begin{align}\label{nmupm}
n_{\mu^\pm}=\frac{g_{\mu^\pm}}{2\pi^2}T^3\left(\frac{m_\mu}{T}\right)^2 K_2(m_\mu/T)=g_{\mu^\pm}\left(\frac{m_\mu T}{2\pi}\right)^{3/2}e^{-{m_\mu}/{T}}\;. 
\end{align}
In the temperature interval we consider, $6<T<3$\,MeV we insert into Eq.\,(\ref{nmuperb}) the particle density Eq.\,(\ref{nmupm}) and entropy density Eq.\,(\ref{entrop}) and obtain
\begin{align}\label{nmuperbF} 
\frac{n_{\mu^\pm}}{n_\mathrm{B}}=\frac{45\;g_{\mu^\pm}}{2\pi^2g^s_\ast}\left(\frac{m_\mu}{2\pi T}\right)^{3/2}e^{-{m_\mu}/{T}}\;1.16\times 10^{10}\;.
\end{align}

\begin{figure}[t]
\begin{center}
\includegraphics[width=11.5cm]{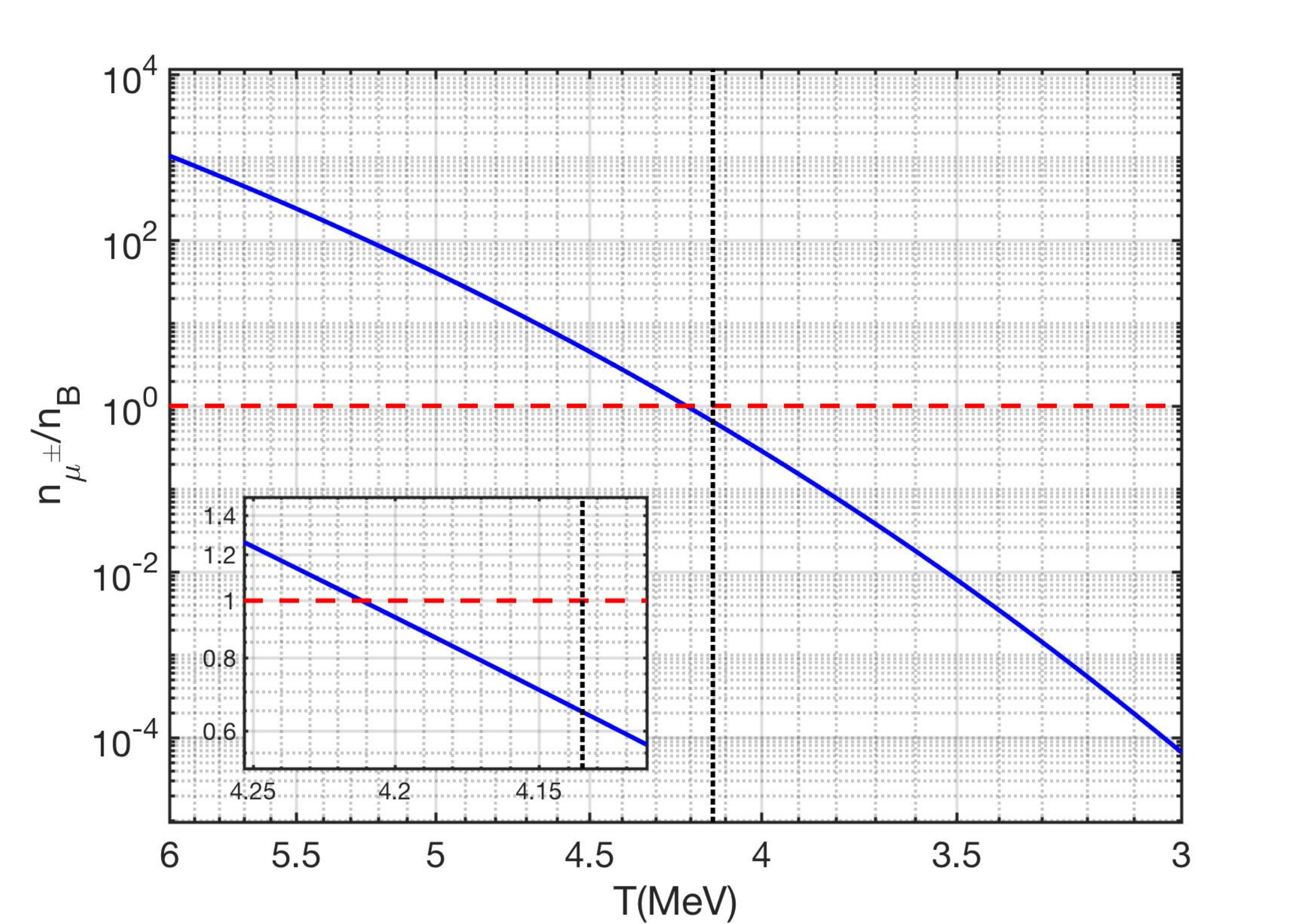}
\caption{\label{DensityRatio_fig}The density ratio between $\mu^\pm$ and baryons as a function of temperature, insert resolves the coincidence of the freeze-out condition of muons (vertical line) occurring near to the cosmic baryon abundance just where the density ratio $n_{\mu^\pm}/n_\mathrm{B}\approx1$ (horizontal line).}
\end{center}
\end{figure}

In Fig.\,\ref{DensityRatio_fig} we show the muon to baryon density ratio Eq.\,(\ref{nmuperbF}) as a function of $T$, and considering the accidental near equality of the equal abundance of baryons and muon pairs we show the detail in the insert: The number density $n_{\mu^\pm}$ and $n_\mathrm{B}$ abundances are equal at the temperature $T\approx4.212\,\mathrm{MeV} \gtrapprox T_\mathrm{disappear}=4.135\,\mathrm{MeV}$, where $n_{\mu^\pm}/n_\mathrm{B}(T_\mathrm{disappear})\approx0.644$. We see that the muon pair abundance $T=6$\,MeV exceeds that of baryons by a factor 1000. 

 \section{Conclusion} \vskip-0.2cm
We compared $\mu^\pm$ production, and decay rates as a function of temperature. The temperature at which $\mu^\pm$ disappear from the Universe is about $T_\mathrm{disappear}\approx 4.135$\,MeV. Below this temperature the $\mu^\pm$ decay rate is faster than the production rate. To characterize the physics situation more precisely, we also evaluated the density ratio between $\mu^\pm$ and baryons in the Universe as a function of temperature.

We presented the thermal production and the natural decay rate of muons $\mu^\pm$ in the primordial Universe as a function of temperature. We concluded that $\mu^\pm$ disappear from the Universe at about $T=T_\mathrm{disappear}\approx 4.135$ MeV, the temperature at which the natural $\mu^\pm$ decay rate overwhelms production rate. The characteristic expansion rate $1/H(T_\mathrm{disappear})= 0.084$\,s in that epoch is $3.8 \times 10^4$ longer compared to the muon lifespan $\tau_\mu=2.2\mu$\,s, therefore muons vanish quasi-instantaneously on the scale of the Universe expansion time.

We considered as a function of temperature the density ratio between $\mu^\pm$ and the baryon inventory in the Universe. We assumed that the ratio $B/S$, baryon per entropy, measured today applies in this primordial epoch. Using this understanding of Universe entropy content, and of $e\bar e$ annihilation neutrino reheating, we have shown that both abundances are equal within the error margin of measured entropy and baryon content values: At the temperature $T\approx4.212$ MeV we have $n_{\mu^\pm}/n_\mathrm{B}=1$ while the density ratio at the muon disappearance temperature $n_{\mu^\pm}/n_\mathrm{B}(T_\mathrm{disappear})\approx0.644$. Since only about half of baryons are protons, as long as muons are present, we have always several charged nonrelativistic muons for each proton and then, rather suddenly, muons disappear.
 
The primary insight of this work is that aside of protons and neutrons, other nonrelativistic charged particles, both positively and negatively charged muons, $\mu^\pm$, are present in kinetic thermal equilibrium and in non-negligible abundance $T>T_\mathrm{disappear}\approx4.1$\,MeV, with a cosmic coincidence of muon pair and baryon abundances coinciding at $T_\mathrm{disappear}$ in regard to baryon abundance in the Universe. Presence of muon pairs offers a new and tantalizing model building opportunity for anyone interested in baryon-antibaryon separation in the primordial Universe, strangelet formation, and perhaps other exotic primordial structure formation mechanisms. Moreover, our result  shows that muons remain in thermal equilibrium abundance in the entire time period in which strangeness evolves down to $T_\mathrm{disappear}\simeq 4.1$\;MeV. We will return to this context in another report~\cite{Rafelski:2020pdi}.\\[-0.3cm]

\begin{figure}[htb]
\centerline{%
\includegraphics[width=5.6cm]{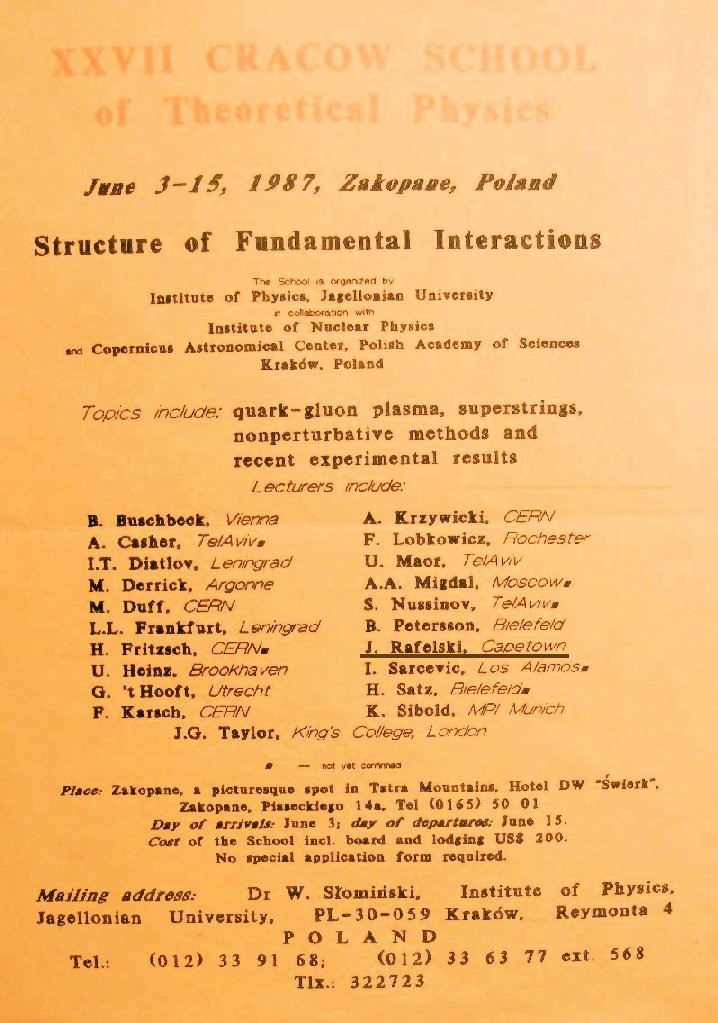}\hspace{0.1cm}
\includegraphics[width=6.7cm]{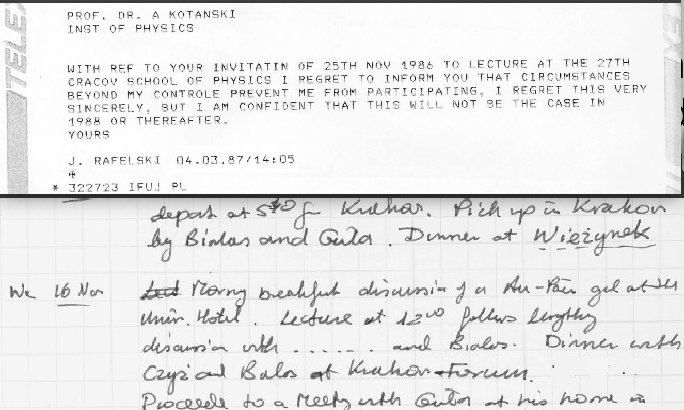}} 
\vskip -8cm\hspace{5.7cm}
\includegraphics[width=6.7cm]{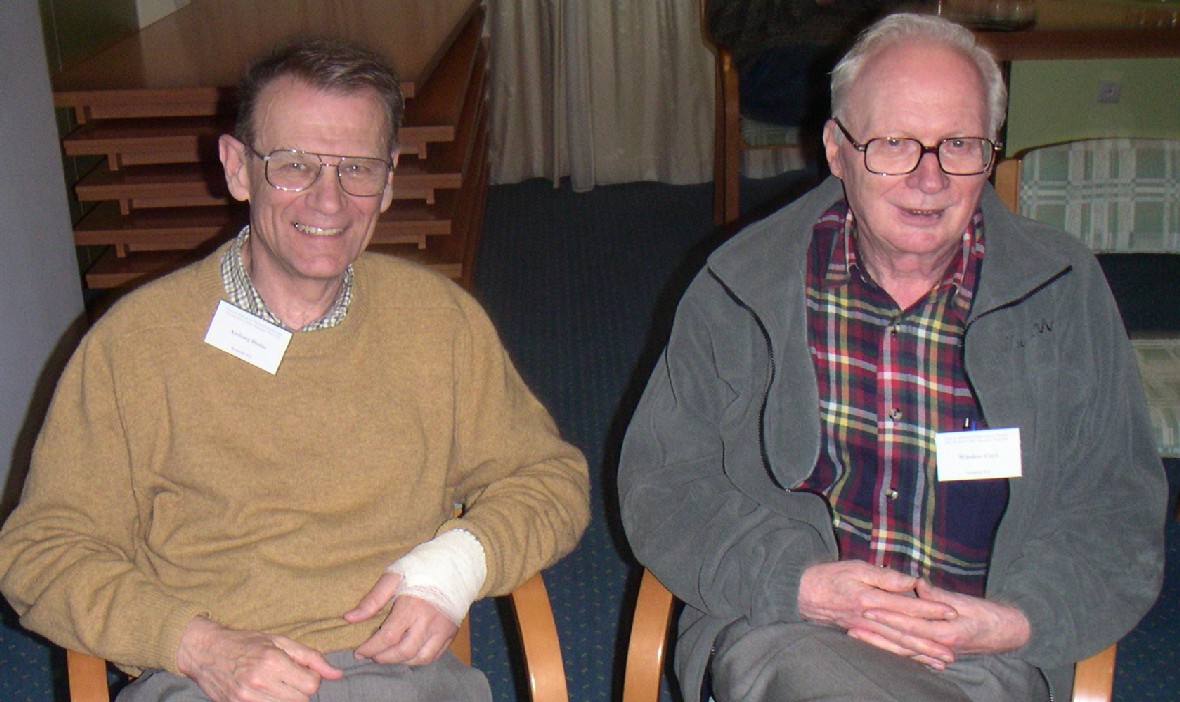}\vskip 4cm
\caption{\label{Fig:87+88}On the left: the poster of the 1987 School; on the right: top image of Andrzej Bialas and Wieslaw Czyz at a Zakopane school about 20 years ago (photo by J. Rafelski); below, Jan's apology of March 1987 regarding inability to attend the School in 1987; bottom right Jan's journal entry addressing events at a year later 16 November 1988 lecture in Krak\.ow}
\end{figure}

{\it Acknowledgements} Jan would like to thank Michal Praszalowicz for the kind invitation to lecture at the jubilee 60th Krak\.ow School of Theoretical Physics, originally set in the splendid context of Krak\.ow and the Tatry Mountains in Zakopane.\\
\centerline{{\large Appendix: {\bf Personal reminiscences related to the Jubilee}}}
\vskip 0.3cm
One of the authors (Jan Rafelski) has a long standing personal and professional connection with the Krak\.ow Summer School of Theoretical Physics: He was born in Krak\.ow and through Andrzej Bialas at CERN more than 40 years ago Jan discovered his research interests of the period overlapped with those of the research groups in his native city; he agreed, see Fig.\,\ref{Fig:87+88} LHS, to lecture at a Summer Zakopane event 1987 - which was more than 20 years after he had illegally left communist Poland.

This first homecoming to Zakopane was set at a time when Jan was professor at the University of Cape Town (UCT), in South Africa, then without consular relations with Poland. At that time there was a proxy war in Angola between the Apartheid regime and a Cuban expedition army; it is said that Polish support units were also present. As it happened, the plan for Jan attending the School in 1987 collapsed in March 1987, see Fig.\,\ref{Fig:87+88} on RHS -- Jan had resigned his position at UCT. Breaking an invitation especially given the personal importance required corrective action; fortunately, an opportunity arose a little more than a year later. 

It is normal not to remember a lecture event for very long. Yet today many details surrounding my November 15-16, 1988 return to Krak\.ow remain sharp in my memory. Seen from perspective of colleagues in Krak\.ow, into a grey winterly landscape of the state of war in Poland, filled with anticipation of a coming change, a colleague from Arizona parachuted in. Jan was received with proverbial Polish hospitality and honors, see Fig.\,\ref{Fig:87+88} RHS bottom. His lecture on Wednesday November 16, 1988 at noon in a traditional Bialas circle was well attended. Many deep friendships with Krak\.ow colleagues followed. 

Looking back one sees that over the last 25 years at least 22  manuscripts were published in Acta Physics Polonica B (including proceedings series) which Jan has coauthored. Of these at least 12 with 300+ published APPB pages address strangeness in high energy nuclear collisions. Furthermore, a joint Tucson-Krak\.ow NATO funded project arose we called SHARE: Statistical Hadronization with Resonances. SHARE has its own 100\rq s printed pages in international research journals.


\end{document}